\documentclass[10pt,letterpaper]{article}
\usepackage{opex3}
\usepackage{geometry}
\usepackage{amsmath}
\usepackage{mathrsfs}
\usepackage{amssymb}
\usepackage{caption}
\usepackage{subcaption}
\usepackage{xfrac}
\usepackage{xcolor}

\DeclareMathOperator{\rect}{rect}

\DeclareMathOperator{\sgn}{sgn}

\begin{document}

\title{\bfseries Rapid Generation of Light Beams Carrying Orbital Angular Momentum}
\author{Mohammad Mirhosseini$^1$, Omar S. Maga\~na-Loaiza$^1$, Changchen Chen$^1$, Brandon Rodenburg$^1$, \mbox{Mehul Malik}$^1$, and Robert W. Boyd$^{1,2}$}
\address{
$^1$The Institute of Optics, University of Rochester. 320 Wilmot BLDG, 275 Hutchison Rd, Rochester NY 14627, USA
\\
$^2$Department of Physics, University of Ottawa, Ottawa, ON K1N 6N5, Canada
}\email{mirhosse@optics.rochester.edu}
\begin{abstract}We report a technique for encoding both amplitude and phase variations onto a laser beam using a single digital micro-mirror device (DMD). Using this technique, we generate Laguerre-Gaussian and vortex  orbital-angular-momentum (OAM) modes, along with modes in a set that is mutually unbiased with respect to the OAM basis. Additionally, we have demonstrated rapid switching among the generated modes at a speed of 4 kHz, which is much faster than the speed regularly achieved by spatial light modulators (SLMs). The dynamic control of both phase and amplitude of a laser beam is an enabling technology for classical communication and quantum key distribution (QKD) systems that employ spatial mode encoding. 
\end{abstract}



\section{Introduction}

Introduced in 1992 by Allen \emph{et al.}  \cite{Allen1992}, orbital angular momentum (OAM) of light has emerged as a useful tool for quantum information science. The discrete unbounded state-space provided by OAM modes have recently been used to increase the channel capacity of a free-space communication link \cite{Wang2012}. Additionally, OAM encoding has been suggested as a prime candidate for realizing multilevel quantum key distribution (QKD) \cite{Boyd:2011wv, Mirhosseini:2013wy}. It has been proposed that the use of a multilevel encoding scheme in such cryptography systems can increase the tolerance to eavesdropping attacks \cite{Cerf:2002fp,Bourennane2002}.

An OAM mode can be created by simply imposing an \(e^{i\ell\phi}\) phase structure onto a laser beam. This task can be achieved by using computer generated holograms \cite{Ando2009, Gibson2004}, q-plates \cite{Marrucci:2006ga}, or spiral phase plates \cite{Sueda:2004wv}. Generation of secure quantum keys in a QKD system, however, requires two or more mutually unbiased bases (MUBs) \cite{Bennett1984}.  Construction of a MUB for the basis of OAM modes often requires both phase and amplitude modulation of a laser beam \cite{Malik:2012ka,OSullivan:2012gj}. Phase-only spatial light modulators (SLMs) have been used previously for creating such modes \cite{Gruneisen2008a}. However, the vast majority of commercially available SLMs are limited by a frame refresh-rate of about 60 Hz which considerably limits the speed of operation of any system based on this technology.

A Digital Micro-mirror Device (DMD) is an amplitude-only spatial light modulator. The high speed, wide range of operational spectral band-width, and high power threshold of a DMD make it a very useful tool for a variety of applications-- from 3D computational imaging \cite{Sun:2013jx} to optical control of neuronal activity \cite{Zhu:2012cu}. Further, variations of DMDs are commercially available for a fraction of the cost of a phase-only SLM. Recently, a DMD was used to encode a varying phase structure onto a beam \cite{Ren:2010dc, Rodrigo:2006jn}. Intensity shaping of spatial modes can be achieved by switching the micro-mirrors on and off rapidly. However, the modes created using this process are not temporally stable and have the desired intensity profile only when averaged by a slow detector. Alternatively, a pseudo-random pixel dithering has been used by previous workers to achieve continuous amplitude modulation \cite{Lerner:2012ee}. This method creates modes that have a qualitatively correct amplitude profile, but the purity of the modes are compromised in this process due to presence of a diffuse speckle pattern resulting from the random structure of such holograms \cite{Cohn:1996}. 

In this paper, we encode phase and amplitude information by modulating the position and the width of a binary amplitude grating, respectively. By realizing such holograms on a DMD, we have successfully created LG modes, OAM vortex modes, and angular (ANG) modes which form a MUB for the OAM basis. Furthermore, we have directly demonstrated active switching of the generated modes at a speed as high as 4 kHz.

\section{Theory}

To introduce our technique, we consider a one-dimensional binary amplitude grating. The transmission function for this grating can be written as
\begin{equation}\label{eq:PulseTrain}
T(x) =  \sum_{m= -\infty}^{\infty} \rect\left[\frac{x-(m+p)x_0}{w x_0}\right],
\end{equation} 
where
\begin{equation}
\rect(u) = \left\{ 
  \begin{array}{l l}
    1 & \quad \text{if}\quad |u| \leq 1/2 , \\
    0 & \quad \text{else}.
  \end{array} \right.
\end{equation} 
This function can be pictured as a pulse train with a period of \(x_0\). The parameters \(p\) and \(w\) are unitless quantities that set the position and the width of each pulse and are equal to constant values for a uniform grating. Here we show that it is possible to locally change the value of these parameters to achieve phase and amplitude modulation of the optical field.

\begin{figure}[b]
\centerline{\includegraphics[width=13cm]{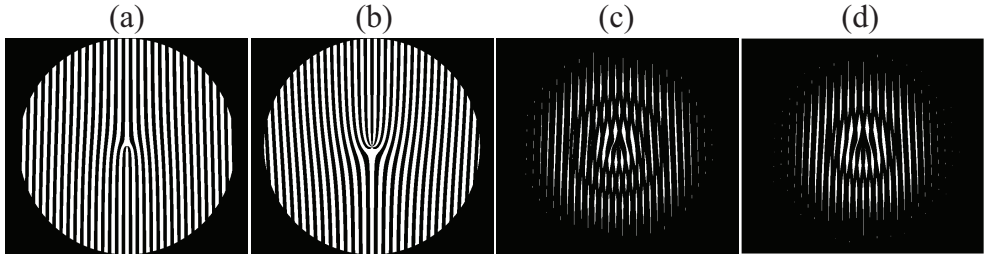}}
\caption{Binary hologram for generating (a) \(\ell= 2\) vortex OAM mode, (b) \(\ell= -5\) vortex OAM mode, (c) \({LG}_{22}\) and (d) \({LG}_{21}\) .}\label{fig:HolOAM}
\end{figure}

 The transmittance function \(T(x)\) is a periodic function, and so it can be expanded as a Fourier series as
\begin{equation}
T(x) =  \sum_{m= -\infty}^{\infty} T_m \exp\left[i2\pi m\left(\frac{x}{x_0}\right)\right].
\end{equation} 
The coefficients \(T_m\) are functions of parameters \(w\) and \(p\) and are given by
\begin{equation}
T_m = \frac{\sin({\pi m w})}{\pi m} e^{i2\pi m p},
\end{equation} 
where we have assumed that both \(w\) and \(p\) are bounded between zero and one. Now, we consider a case where \(p(x)\) and \(w(x)\) are functions of \(x\) and the binary grating is illuminated by a monochromatic plane wave. The first-order diffracted light can be written as
\begin{equation}\label{eq:T1}
T_1(x) = \frac{1}{\pi}\sin[{\pi w(x)}] e^{i2\pi p(x)}.
\end{equation} 
It can be seen that \(w(x)\) is related to the amplitude of the diffracted light while \(p(x)\) sets its phase. Therefore, the phase and the amplitude of the diffracted light can be controlled by setting the parameters \(p(x)\) and \(w(x)\). In the language of communication theory, these methods are sometimes referred to as pulse position modulation (PPM) and pulse width modulation (PWM) \cite{Brown:1969ui,Lee:1979wp, Davis:2003}. Eq. \ref{eq:T1} provides a good approximation for slowly varying \(p(x)\) and \(w(x)\) functions \cite{KIRK:1971kc,Chu:1972vd}, when a mixed Fourier-Taylor analysis can be used to derive this formula \cite{Davis:1999ku}. 

The analysis above treats a one-dimensional case. We have generated a two-dimensional grating by thresholding a rapidly varying modulated carrier as
\begin{equation}\label{eq:modulated}
T(x,y) = 1/2 +1/2\sgn{ \left({\cos\left[2\pi x/x_0 + \pi p(x,y)\right]-\cos[\pi w(x,y)]} \right)}.
\end{equation} 
Here, \(\sgn(x,y)\) is the sign function. It is easy to check that in the limit where \(w(x,y)\) and \(p(x,y)\) are slowly varying, this formula reproduces the pulse train described above.
We can find the corresponding \(w(x,y)\) and \(p(x,y)\) functions for a general complex scalar field \(\mathcal{A}(x,y)e^{i\varphi(x,y)}\) according to the relations
\begin{equation}
w(x,y)= \frac{1}{\pi} \arcsin[\mathcal{A}(x,y)],
\end{equation} 
\begin{equation}
p(x,y)= \frac{1}{\pi}\varphi(x,y).
\end{equation} 
We have assumed that the field contains no singularity and thus its amplitude can be normalized to have a maximum of unity.

We have designed two-dimensional binary amplitude holograms to generate Laguerre-Gaussian (LG) modes. Fig. \ref{fig:HolOAM} shows sample holograms designed for generation of vortex OAM and LG modes. It can be seen that in both cases, the holograms have the familiar forked structure. The gratings designed for vortex modes have a fairly uniform width across the aperture whereas for the case of LG modes the gratings gradually disappear where the amplitude gets negligibly small.

\section{Experiment}
 
A digital micro-mirror device (DMD) is an amplitude-only spatial light modulator \cite{Hornbeck}. The device consists of a series of micro-mirrors that can be controlled in a binary fashion by setting the deflection angle of each individual mirror to either +12  or \(-12\) degrees.
This enables the on-demand realization of binary gratings that can be switched at very high speeds using an external digital signal \cite{Dudley:2003kr}. 

\begin{figure*}[b]
\centerline{\includegraphics[width=9.5cm]{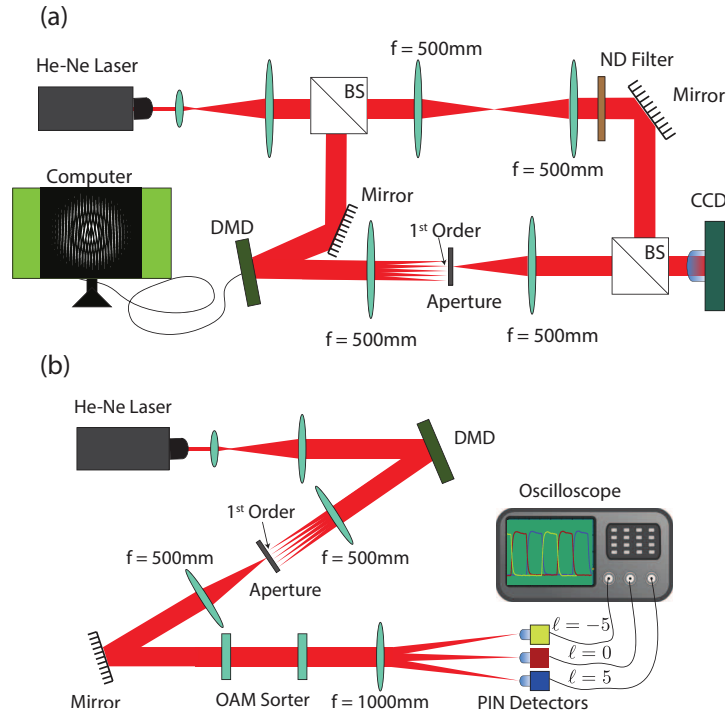}}
\caption{Schematic diagram of the experimental setup for a) measuring intensity profiles and phase interferograms. b) switching among three OAM modes and detecting them in real time.}\label{fig:Setup}
\end{figure*}

\begin{figure*}[t]
\centerline{\includegraphics[width=13cm]{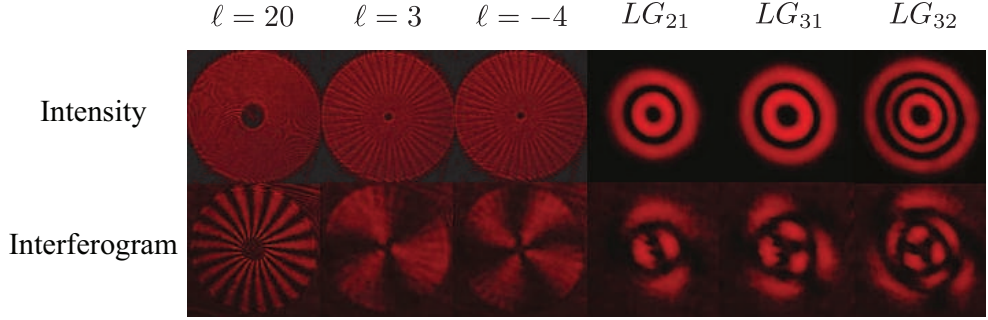}}
\caption{Intensity and interferograms of three vortex modes (to the left) and LG modes (to the right). The interferograms demonstrate the phase structure of the beams and have been obtained by interfering the generated modes with a plane wave. }\label{fig:Res1}
\end{figure*}

\begin{figure*}[b]
\centerline{\includegraphics[width=11cm]{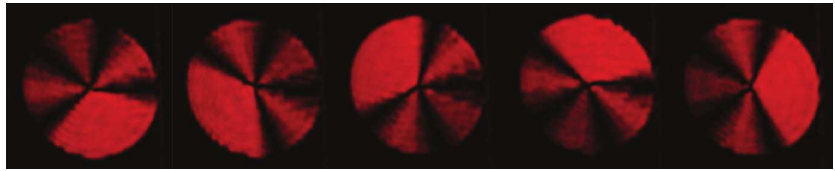}}
\caption{Intensity patterns for ANG modes constructed from superposition of vortex OAM modes from the set \(\ell \in [-2:2]\).}\label{fig:Res3}
\end{figure*}

In our experiment, we generate spatial modes by loading computer generated holograms described in the previous section onto a DMD. We have used a Texas Instrument DLP3000 DMD for this task. This device has a resolution of 608 \(\times\) 684, a micro-mirror size of 7.5 \(\mu\)m, and an array diffraction efficiency of 86 \%. The holograms for generating modes are created by modulating a grating function with 20 micro-mirrors per each period. As shown in  Fig. \ref{fig:Setup}, a 4f imaging system along with an aperture separates the first order diffracted light. We use a charge-coupled-device (CCD) camera for measuring the intensity profile of the generated modes. In addition, we have used a Mach-Zehnder interferometer for verifying the phase patterns of the created beams. A collimated plane-wave reference from the same laser is interfered with the modes generated by the DMD to obtains interferograms. 

We have created OAM modes and LG modes using the setup. Fig. \ref{fig:Res1} shows the intensity pattern along with the interferogram for a few of these modes. The structure of the interferograms is in agreement with the helical phase pattern of OAM modes, with the number of azimuthal dark lines equating to the OAM quantum number in each case. The LG modes have the same azimuthal phase structure as the vortex modes. In addition, these modes contain \(p+1\) rings in the radial direction where \(p\) is the radial quantum number \cite{Allen1992}. The change of sign of the optical field between each two consecutive rings is an intrinsic property of the LG modes and it can be clearly noticed in the interferograms.

\begin{figure*}[t]
\centerline{\includegraphics[width=13cm]{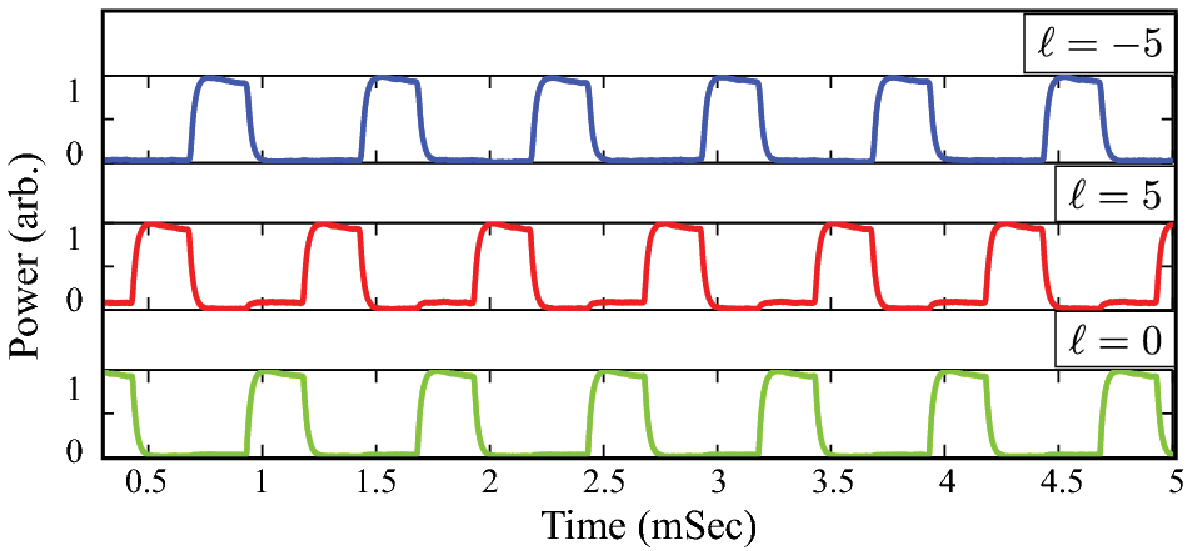}}
\caption{Detected power as function of time for three vortex OAM modes of \(\ell = 5\), \(\ell = -5\) and \(\ell = 0\). It is seen that the generated mode can be changed on a time-scale of 0.25 ms (measured at FWHM) corresponding to a switching rate of 4 kHz. The detecter power is divided to its maximum value in each case.}\label{fig:Res2}
\end{figure*}

The examples above demonstrate the possibility of using binary holograms to coherently control both phase and amplitude of a beam. A low number of pixels per each period of the binary grating results in quantization errors in encoding phase and intensity. On the other hand, the total number of grating periods within the incident beam on the DMD sets an upper limit on the spatial bandwidth of the generated modes. Consequently, a large total number of micro-mirrors is preferable for generating high quality modes. This can be achieved by using newer generations of DMDs (for example TI DLP 9500) that are available with HD resolution (1920 \(\times\) 1080, which equals the highest resolution available for phase-only SLMs).

 Another set of modes that are needed for OAM-based quantum key distribution (QKD) is the set of angular (ANG) modes. These modes form a mutually unbiased basis (MUB) with respect to the OAM basis \cite{Malik:2012ka}. Mathematically, these modes can be described as

\begin{equation}
\label{eq:ANGMode}
\theta_{j,N_\ell}(r,\varphi) = \frac{1}{\sqrt{2N_\ell+1}} \sum_{\ell=-L}^{L} u_{\ell}(r,\varphi) e^{-i 2 \pi j\ell/(2N_\ell+1)}.
\end{equation}

Here, \(u_{\ell}(r,\varphi)\) represents each OAM mode (either LG or vortex) and \(N\) is the total number of them. The simultaneous use of both OAM and ANG modes is necessary for achieving security in an OAM-based QKD system \cite{Malik:2012ka}. We have generated ANG modes using our technique. The amplitude profile of a number of these modes is presented in Fig. \ref{fig:Res3}.
 
We have measured the efficiency of generating OAM modes to be about \(1.5\%\). This number can be increased be optimizing the angle of incidence of the beam on the mirror array. However, the maximum theoretical efficiency of a binary amplitude grating is about \(10 \%\) \cite{GoodmanBook}. This is much smaller than the mode generation efficiency of phase-only SLMs, which are typically above \(50\%\), and can even reach values as high as \(90\%\). Nonetheless, our main main motivation for using the presented technique is free-space QKD with faint laser pulses, which typically requires an average value of 0.1 photon per pulse in order to be secure against number splitting attacks \cite{Gisin2002}. Taking this requirement into account, the low efficiency of a DMD in the generation of spatial modes is not a limitation but may even be considered advantageous. Due to its wedge-like structure, the efficiency of generating a \( \theta_{j,N_\ell}(r,\varphi) \) ANG mode is roughly equal to \(\frac{1}{2N_\ell+1}\) times the efficiency of generating an OAM mode. Within the experimental errors, we have verified this in property in our experiment.
 
Using a DMD for generating OAM modes gives us the ability to switch between different modes at very high speeds. This method involves a much smaller number of optical elements as compared to the conventional techniques where OAM modes are generated using a series of separated forked holograms and are multiplexed using beam splitters \cite{Wang2012}. As a direct test of the high-speed capability of our mode-generation system, we have implemented dynamic switching among vortex OAM modes with quantum number \(\ell = 5, -5\) and 0. The computer generated holograms for these modes were loaded onto the memory of the DMD and the switching was achieved by using a clock signal. We have used the mode sorter described in \cite{Lavery2012,Mirhosseini:2013wy} to map the input modes to a series of separated spots. We then measured the intensity corresponding to each mode using a high bandwidth PIN detector at the positions corresponding to each mode. Fig. \ref{fig:Res2} demonstrates the measured values. It can be seen that the OAM quantum number of the generated modes can be controlled at a speed of 4 kHz. This is a clear demonstration of the capability to rapidly switch among such modes. It should be noted that the speed reported above is a limitation imposed by the DMD we have used and it is not inherent to this technique. In fact, commercially available DMDs can achieve a speed of as high as 32 kHz (TI DLP7000). Further, these devices are available for a fraction of the cost of a phase-only spatial light modulator.

\section{Conclusions}
In this work, we have introduced a method for generation of an arbitrary complex scalar field using a digital micro-mirror device (DMD). We have used a single binary amplitude-only hologram to encode both amplitude and phase information onto a beam. With this technique, we have generated LG modes, OAM vortex modes, and angular (ANG) modes. The ANG basis is mutually unbiased with respect to the OAM basis and is very important for OAM-based QKD systems. In addition, we have demonstrated the generation and active switching of OAM modes at a speed of 4 kHz by dynamically changing the hologram realized by the DMD. This technology has potential applications in classical communication systems and QKD systems that are based on OAM encoding. 
\section{Acknowledgements}
We gratefully acknowledge valuable discussions with M. O'Sullivan, Z. Shi, E. Karmi, J. Vornehm, and D. Gauthier. This work was supported by the DARPA/DSO InPho program and the Canadian Excellence Research Chair (CERC) program and the European Commission through a Marie Curie fellowship. OSML also acknowledges support from the CONACyT.

\end{document}